\documentclass[12pt,thmsa]{article}

\begin{document}

\textbf{Relativistic\ Electrodynamics without Reference Frames. }

\textbf{Clifford Algebra Formulation }\bigskip

\qquad Tomislav Ivezi\'{c}

\qquad \textit{Ru%
\mbox
 {\it{d}\hspace{-.15em}\rule[1.25ex]{.2em}{.04ex}\hspace{-.05em}}er Bo\v
{s}kovi\'{c} Institute, P.O.B. 180, 10002 Zagreb, Croatia}

\textit{\qquad ivezic@irb.hr\bigskip }

In the usual Clifford algebra formulation of electrodynamics the Faraday
bivector field $F$ is expressed in terms of \emph{the observer dependent}
relative vectors $\mathbf{E}$ and $\mathbf{B.}$ In this paper we present
\emph{the observer independent }decomposition of $F$ by using the vectors
(grade-1) of electric $E$ and magnetic $B$ fields and we develop the
formulation of relativistic electrodynamics which is independent of the
reference frame and of the chosen coordinatization. We also present
equivalent formulations by using the multivector (Clifford aggregate) $\Psi
=E-e_{5}cB$ (for the real Clifford algebra; $e_{5}$ is the (grade-4)
pseudoscalar) or the complex 1-vector $\Psi =E-icB$ (for the complex
Clifford algebra; $i$ is the unit imaginary). In such formulations the
Maxwell equations become a Dirac like relativistic wave equation for the
free photon. \bigskip

\noindent PACS numbers: 03.30.+p \bigskip

In the usual Clifford algebra treatments, e.g. $\left[ 1-3\right] $, of
electrodynamics the Maxwell equations (ME) are written as a single equation
using the electromagnetic field strength $F$ (a bivector) and the gradient
operator $\partial $ (1-vector). In order to get the more familiar form $F$
is expressed in terms of relative vectors $\mathbf{E}$ and $\mathbf{B}$ by
making a space-time split, which depends on \emph{the observer velocity. }%
(For the Clifford algebra see also the very stimulating modern textbook $%
\left[ 4\right] $.)

In the recent works $\left[ 5,6\right] $ an invariant formulation of special
relativity (SR)\ is proposed (see also $\left[ 7\right] $) and compared with
different experiments $\left[ 6\right] $. There a physical quantity in the
four-dimensional (4D) spacetime is mathematically represented either by a
true tensor (when no basis has been introduced) or equivalently by a
coordinate-based geometric quantity (CBGQ) comprising both components and a
basis (when some basis has been introduced). \emph{Such formulation is
independent of the reference frame and of the chosen coordinatization. }It
is compared with the usual covariant formulation, which mainly deals with
the basis components of tensors in a specific, i.e., Einstein's (''E'')
coordinatization. In the ''E'' coordinatization the Einstein synchronization
$\left[ 8\right] $ of distant clocks and Cartesian space coordinates $x^{i}$
are used in the chosen inertial frame of reference (IFR). It is also found
in $\left[ 5\right] $ that the usual transformations of the three-vectors of
the electric and magnetic fields $\mathbf{E}$ and $\mathbf{B}$ are not
relativistically correct.

In this paper it is shown that the space-time split is not relativistically
correct procedure and that the relative vectors are not well-defined
quantities from the SR viewpoint. Further we write the Lorentz
transformations (LT) in a coordinatization independent way. The expressions
for the observer independent stress-energy vector $T(v),$ energy density $U$
(scalar) and the Poynting vector (1-vector) $S$ and the Lorentz force $K$
are directly derived from the ME. Then we present \emph{the observer
independent decomposition of }$F$\emph{\ in terms of 1-vectors }$E$\emph{\
and }$B.$ The equivalent formulations of electrodynamics with $F$, $E$ and $%
B $, the real multivector $\Psi =E-e_{5}cB$ and with the complex 1-vector $%
\Psi =E-icB$ are developed and presented here. It is also shown that in the
real and the complex $\Psi $ formulations the ME become a Dirac like
relativistic wave equation for the free photon. The main idea of the whole
approach is the same as for the invariant SR with true tensors $\left[
5,6\right] $, i.e., that \emph{the physical meaning is attributed, both
theoretically and experimentally, only to the 4D quantities. }

\textit{Brief summary of geometric algebra }$\left[ 1-4\right] $\textit{.}-
We write Clifford vectors in lower case ($a$) and general multivectors in
upper case ($A$). The homogeneous multivectors (of a single grade $r$) are
written $A_{r}.$ The geometric (Clifford) product is written by simply
juxtaposing multivectors $AB$. The degree projection $\left\langle
A\right\rangle _{r}$ selects from the multivector $A$ its $r-$ vector part ($%
0=$ scalar, $1=$ vector, $2=$ bivector, ....). We write the scalar simply as
$\left\langle A\right\rangle .$ The geometric product of $A_{r}$ with $B_{s}$
decomposes into $A_{r}B_{s}=\left\langle AB\right\rangle _{\
r+s}+\left\langle AB\right\rangle _{\ r+s-2}...+\left\langle AB\right\rangle
_{\ \left| r-s\right| }.$ The inner and outer products are the lowest-grade
and the highest-grade terms respectively of the above series $A_{r}\cdot
B_{s}\equiv \left\langle AB\right\rangle _{\ \left| r-s\right| }$ and $%
A_{r}\wedge B_{s}\equiv \left\langle AB\right\rangle _{\ r+s}.$ For vectors $%
a$ and $b$ we have $ab=a\cdot b+a\wedge b,$ and $a\cdot b\equiv (1/2)(ab+ba)$%
, $a\wedge b\equiv (1/2)(ab-ba).$ Reversion is defined by $\widetilde{AB}=%
\widetilde{B}\widetilde{A},$ $\widetilde{a}=a,$ for any vector $a$, and it
reverses the order of vectors in any given expression. The operation called
the complex reversion is introduced for the complex Clifford numbers and
denoted by an overbar $\overline{\Psi .}$ It takes the complex conjugate of
the scalar (complex) coefficient of each of the 16 elements in the algebra,
and reverses the order of multiplication of vectors in each multivector. In
the treatments, e.g., $\left[ 1-4\right] $, one usually introduces the
standard basis. The generators of the spacetime algebra (STA) are taken to
be four basis vectors $\left\{ \gamma _{\mu }\right\} ,\mu =0...3,$
satisfying $\gamma _{\mu }\cdot \gamma _{\nu }=\eta _{\mu \nu }=diag(+---).$
The $\gamma _{\mu }$ generate by multiplication a complete basis, the
standard basis, for STA: $1,\gamma _{\mu },\gamma _{\mu }\wedge \gamma _{\nu
},\gamma _{\mu }\gamma _{5,}\gamma _{5}$ ($16$ independent elements). $%
\gamma _{5}$ is the pseudoscalar for the frame $\left\{ \gamma _{\mu
}\right\} .$

In $\left[ 1-3\right] $ one introduces a space-time split and the relative
vectors. By singling out a particular time-like direction $\gamma _{0}$ we
can get a unique mapping of spacetime into the even subalgebra of STA. For
each event $x$ this mapping is specified by $x\gamma _{0}=ct+\mathbf{x,\quad
}ct=x\cdot \gamma _{0},\ \mathbf{x}=x\wedge \gamma _{0}.$ The set of all
position vectors $\mathbf{x}$ is the 3D position space of the observer $%
\gamma _{0}$ and it is designated by $P^{3}$. The elements of $P^{3}$ are
called \textit{the relative vectors} (relative to $\gamma _{0})$ and they
will be designated in boldface.

The explicit appearance of $\gamma _{0}$ implies that the space-time split
is observer dependent. If we consider the position 1-vector $x$ in another
relatively moving IFR $S^{\prime }$ (characterized by $\gamma _{0}^{\prime }$%
) then the space-time split in $S^{\prime }$ and in the ''E''
coordinatization is $x\gamma _{0}^{\prime }=ct^{\prime }+\mathbf{x}^{\prime
} $\textbf{.} This $x\gamma _{0}^{\prime }$ is not obtained by the LT from $%
x\gamma _{0}.$ (The hypersurface $t^{\prime }=const.$ is not connected in
any way with the hypersurface $t=const.$) Thus the spatial and the temporal
components ($\mathbf{x}^{\prime }$, $t^{\prime }$) of some geometric 4D
quantity ($x$) are not physically well-defined quantities. Only their union
is physically well-defined quantity. It has to be noted that in SR different
coordinatizations of an IFR are allowed and they are all equivalent in the
description of physical phenomena, see, e.g., $\left[ 5\right] .$ Thence
instead of the standard basis $\left\{ \gamma _{\mu }\right\} $ (the ''E''
coordinatization) we can use some basis $\left\{ e_{\mu }\right\} $ (the
metric tensor of $M^{4}$ is defined as $g_{\mu \nu }=e_{\mu }\cdot e_{\nu }$%
). Then, e.g., $x$ can be decomposed in the $S$ and $S^{\prime }$ frames and
in the standard basis $\left\{ \gamma _{\mu }\right\} $ and some
non-standard basis $\left\{ e_{\mu }\right\} $ as $x=x^{\mu }\gamma _{\mu
}=..=x^{\mu ^{\prime }}e_{\mu ^{\prime }}.$ The primed quantities are the
Lorentz transforms of the unprimed ones.

\textit{Lorentz transformations}.-\textbf{\ }In, e.g., $\left[ 1-3\right] $,
the LT are considered as active transformations and they are described with
rotors $R,$ $R\widetilde{R}=1,$ in the usual way as $p\rightarrow p^{\prime
}=Rp\widetilde{R}=p_{\mu ^{\prime }}\gamma ^{\mu }.$ Every rotor in
spacetime can be written in terms of a bivector as $R=e^{\theta /2},$ $%
\theta =\alpha \gamma _{0}n,$ $\tanh \alpha =\beta $, $\beta $ is the scalar
velocity in units of $c$, and $n$ is any unit space-like vector orthogonal
to $\gamma _{0}$.

Here a coordinatization independent form for the LT is introduced and it can
be interpreted both in an active or a passive way. We introduce 1-vector $%
u=cn,$ which represents the proper velocity of the frame $S$ \emph{with
respect to itself}. Taking that $v$ is 1-vector of the velocity of $%
S^{\prime }$ relative to $S$ we write the component form of $L$ in some
basis $\left\{ e_{\mu }\right\} $ as
\begin{equation}
L_{\nu }^{\mu }=g_{\nu }^{\mu }+2u^{\mu }v_{\nu }c^{-2}-(u^{\mu }+v^{\mu
})(u_{\nu }+v_{\nu })/c^{2}(1+u\cdot v/c^{2}),  \label{L}
\end{equation}
or as a CBGQ $L=L_{\nu }^{\mu }e_{\mu }e^{\nu }.$ $R$ connected with such $L$
is $R=L/(\widetilde{L}L)^{1/2}=L_{\nu }^{\mu }e_{\mu }e^{\nu }/(\widetilde{L}%
L)^{1/2},$ where $\widetilde{L}L=8(\gamma +1)$ and $\gamma =u\cdot v/c^{2}.$
It can be also written as $R=\exp ((\alpha /2)(u\wedge v)/\left| u\wedge
v\right| )$. One can solve $L_{\nu }^{\mu }$ in terms of $R$ as $L_{\nu
}^{\mu }=\left\langle e_{\nu }\widetilde{R}e^{\mu }R\right\rangle .$ The
usual results are recovered when the standard basis $\left\{ \gamma _{\mu
}\right\} $ is used. But our results for $L$ and $R$ hold also for other
coordinatizations.

\textit{The} $F$ \textit{formulation}.- We start the exposition of
electrodynamics writing ME in terms of $F$ $\left[ 1-3\right] $. The source
of the field is the electromagnetic current $j$ (1-vector field). Then using
$\partial $ (1-vector) ME can be written as a single equation
\begin{equation}
\partial F=j/\varepsilon _{0}c,\quad \partial \cdot F+\partial \wedge
F=j/\varepsilon _{0}c.  \label{MEF}
\end{equation}
The equation (\ref{MEF}) encodes all of the Maxwell equations. For the given
$j$ the Clifford algebra formalism enables one to find $F.$ Namely $\partial
$ is invertible and (\ref{MEF}) can be solved for $F=\partial
^{-1}(j/\varepsilon _{0}c).$ In the Clifford algebra formalism one can
derive the expressions for the stress-energy vector $T(v)$ and the Lorentz
force $K$ directly from ME. One finds from ME, $\left[ 1-3\right] $,

\begin{equation}
T(\partial )=(-\varepsilon _{0}/2)(F\partial F)=j\cdot F/c=-K,  \label{TEF}
\end{equation}
where in $(F\partial F)$ the derivative $\partial $ operates to the left and
to the right by the chain rule. $T(v)$ $\left[ 1-3\right] $ is defined in
the $F$ formulation as $T(v)=-(\varepsilon _{0}/2c)\left\langle
FvF\right\rangle _{1}.$ It is a vector-valued linear function on the tangent
space at each spacetime point $x$ describing the flow of energy-momentum
through a surface with normal $n=n(x);$ $v=cn.$ The r.h.s. of (\ref{TEF})
yields the Lorentz force $K=F\cdot j/c$, or for a charge $q$, $K=(q/c)F\cdot
u,$ where $u$ is the velocity 1-vector of a charge $q$ (it is defined to be
the tangent to its world line).

We write $T(v)$ as a sum of two terms
\begin{eqnarray}
T(v) &=&-(\varepsilon _{0}/2c)\left[ -(F\cdot F)+(2/c^{2})(F\wedge
v)^{2}\right] v+  \nonumber \\
&&-(\varepsilon _{0}/c)\left[ (F\cdot v)\cdot F-(1/c^{2})(F\cdot
v)^{2}v\right] .  \label{ste}
\end{eqnarray}
The first term in (\ref{ste}) is $v$-parallel part ($v-\parallel $) and it
yields the observer independent energy density $U$ ($U=v\cdot T(v)/c$,
scalar) and the second term is $v$-orthogonal part ($v-\perp $) which is $%
(1/c)S$, where $S$ is the Poynting vector (1-vector), and it holds that $%
v\cdot S=0$. From (\ref{MEF}) one can also derive the local (and global)
charge and energy-momentum conservation laws, but it will not be done here.

In $\left[ 1-3\right] $ $F$ is expressed in terms of a relative vector $%
\mathbf{E}$ and a relative bivector $\gamma _{5}\mathbf{B}$ by making a
space-time split in the $\gamma _{0}$ frame $F=\mathbf{E}+\gamma _{5}c%
\mathbf{B,}$ where $\mathbf{E}=(F\cdot \gamma _{0})\gamma _{0}$ and $\gamma
_{5}\mathbf{B}=(F\wedge \gamma _{0})\gamma _{0}$. It can be shown (as
discussed for the space-time split of $x$) that such decomposition of $F$ is
not relativistically correct and that the usual transformations of the
relative vectors $\mathbf{E}$ and $\mathbf{B}$ are not the LT of some
well-defined quantities on the 4D spacetime. This will be presented in
detail in an enlarged version.

\textit{The} $E,$ $B$ \textit{formulation.}- Here we present \emph{the
observer independent} decomposition of $F$ by using 1-vectors $E$ and $B$.
We define
\begin{equation}
F=(1/c)E\wedge v+e_{5}B\cdot v,\quad E=(1/c)F\cdot v,\
e_{5}B=(1/c^{2})F\wedge v,  \label{myF}
\end{equation}
and it holds that $E\cdot v=B\cdot v=0$. $v$ is the velocity (1-vector) of a
family of observers who measures $E$ and $B$ fields. The relation (\ref{myF}%
) establishes the equivalence of the $F$ and $E,$ $B$ formulations. The ME (%
\ref{MEF}) become
\begin{equation}
\partial ((1/c)E\wedge v+e_{5}B\cdot v)=j/\varepsilon _{0}c.  \label{deb}
\end{equation}
In some basis $\left\{ e_{\mu }\right\} $ the ME (\ref{deb}) are
\begin{eqnarray}
\partial _{\alpha }(\delta _{\quad \mu \nu }^{\alpha \beta }v^{\mu }E^{\nu
}-\varepsilon ^{\alpha \beta \mu \nu }v_{\mu }cB_{\nu })e_{\beta }
&=&-(j^{\beta }/\varepsilon _{0})e_{\beta },  \nonumber \\
\partial _{\alpha }(\delta _{\quad \mu \nu }^{\alpha \beta }v^{\mu }cB^{\nu
}+\varepsilon ^{\alpha \beta \mu \nu }v_{\mu }E_{\nu })e_{5}e_{\beta } &=&0,
\label{maeb}
\end{eqnarray}
where $E^{\alpha }$ and $B^{\alpha }$ are the basis components of $E$ and $B$%
, and $\delta _{\quad \mu \nu }^{\alpha \beta }=\delta _{\,\,\mu }^{\alpha
}\delta _{\,\,\nu }^{\beta }-\delta _{\,\,\nu }^{\alpha }\delta _{\,\mu
}^{\beta }.$ We remark that (\ref{maeb}) follows from (\ref{deb}) for
coordinatizations with constant $e_{\mu }$, e.g., for the standard basis $%
\left\{ \gamma _{\mu }\right\} $ (the ''E'' coordinatization).

The comparison of this geometric approach with $E$ and $B$ and the usual
approach with the 3-vectors $\mathbf{E}$ and $\mathbf{B}$ (not relative
vectors) is possible in the ''E'' coordinatization and in an IFR $\mathcal{R}
$ in which the observers who measure $E^{\alpha }$ and $B^{\alpha }$ are at
rest, i.e., $v^{\alpha }=(c,\mathbf{0})$. Then it holds that $E^{0}=B^{0}=0$%
. Note that we can select a particular - but otherwise arbitrary - IFR as
the $\mathcal{R}$ frame, to which we shall refer as the frame of our
''fiducial'' observers (see $\left[ 9\right] $). In $\mathcal{R}$ one can
derive from the ME in the $\left\{ \gamma _{\mu }\right\} $ basis (\ref{maeb}%
) the ME which contain only the space parts $E^{i}$ and $B^{i}$ of $%
E^{\alpha }$ and $B^{\alpha }$, e.g., from the first ME in (\ref{maeb}) one
easily finds $\partial _{i}E^{i}=j^{0}/\varepsilon _{0}c$, etc. Thus in%
\textit{\ }$\mathcal{R}$ the ME (\ref{maeb}) are of the same form as the
usual ME with the 3D $\mathbf{E}$ and $\mathbf{B}$. We notice that observers
with $v^{\alpha }=(c,\mathbf{0})$ are usually considered in the conventional
formulation with the 3D $\mathbf{E}$ and $\mathbf{B.}$ However in relatively
moving $S^{\prime }$ these observers are not at rest whence in $S^{\prime }$
one cannot obtain the usual ME with the 3D $\mathbf{E}^{\prime }$ and $%
\mathbf{B}^{\prime }$ from the transformed ME (\ref{maeb}) with $E^{\alpha
^{\prime }}$ and $B^{\alpha ^{\prime }}$. The dependence of the ME (\ref
{maeb}) on $v$ reflects the arbitrariness in the selection of frame $%
\mathcal{R}$ but at the same time it makes the equations (\ref{maeb})
independent of that choice. The frame $\mathcal{R}$ can be selected at our
disposal, which proves that we don't have a kind of the ''preferred'' frame
theory.

Using (\ref{myF}) we write $T(v)$ as a sum of the $v-\parallel $ and the $%
v-\perp $ parts ($v$ is again the velocity of observers who measure $E$ and $%
B$)
\begin{equation}
T(v)=(-\varepsilon _{0}/2c)(E^{2}+c^{2}B^{2})v+\varepsilon _{0}e_{5}\left[
\left( E\wedge B\right) \wedge v\right] .  \label{tv}
\end{equation}
Then $U=(-\varepsilon _{0}/2)(E^{2}+c^{2}B^{2})$ and $S=c\varepsilon
_{0}e_{5}\left[ \left( E\wedge B\right) \wedge v\right] .$ All these
quantities can be written in some basis $\left\{ e_{\mu }\right\} $ as a
CBGQs, but we will not do it here. One can compare these expressions with
familiar ones from the 3D space considering our definitions in the standard
basis $\left\{ \gamma _{\mu }\right\} $ and in the $\mathcal{R}$ frame. Then
the invariant $U$ and $S$ take the familiar forms $U=(-\varepsilon
_{0}/2)(E^{i}E_{i}+c^{2}B^{i}B_{i})\ $and $S=\varepsilon
_{0}c^{2}\varepsilon _{0\ jk}^{\ i}E^{j}B^{k}\gamma _{i},\quad i,j,k=1,2,3.$

In $\left[ 1-3\right] $ one makes the space-time split and writes $K$ in the
Pauli algebra of $\gamma _{0}$. Since this procedure is observer dependent
we express $K$ in an observer independent way using $E$ and $B$ as $%
K=(q/c)\left[ (1/c)E\wedge v+e_{5}B\cdot v\right] \cdot u.$ In the general
case when charge and observer have distinct worldlines $K$ can be written as
a sum of the $v-\perp $ part $K_{\perp }$ and the $v-\parallel $ part $%
K_{\parallel },$ $K=K_{\perp }+K_{\parallel }.$
\begin{equation}
K_{\perp }=(q/c^{2})(v\cdot u)E+(q/c)(e_{5}B\cdot v)\cdot u,\quad
K_{\parallel }=(-q/c^{2})(E\cdot u)v.  \label{Kaok}
\end{equation}
This is an observer independent decomposition of $K$ and it holds that $%
K_{\perp }\cdot v=0$, $K_{\parallel }\wedge v=0.$ Speaking in terms of the
prerelativistic notions one can say that in the approach with $E$ and $B$ $%
K_{\perp }$ plays the role of the usual Lorentz force lying on the 3D
hypersurface orthogonal to $v$, while $K_{\parallel }$ is related to the
work done by the field on the charge. However \emph{in our formulation of SR
only both components taken together do have physical meaning and they define
}$K$ \emph{both in the theory and in experiments. }When \emph{the complete} $%
K$ is known we can solve the equation of motion, Newton's second law,
written as $K=m(u\cdot \partial )u,$ where $u\cdot \partial $ is the
directional derivative and $(u\cdot \partial )u$ defines the acceleration of
the particle. One can compare these expressions with familiar ones from the
3D space considering our results in $\left\{ \gamma _{\mu }\right\} $ basis
and in the $\mathcal{R}$ frame. Then the whole $K$ does have $K^{i}=K_{\perp
}^{i},$ $i=1,2,3$ which is equal to the usual 3D expression for the Lorentz
force, while $K^{0}=K_{\parallel }^{0},$ and represents the 3D expression
for the work done by the field on the charge.

\textit{The real} $\Psi =E-e_{5}cB$ \textit{formulation}.- Now we consider
the formulation of electrodynamics with $\Psi $ and $\widetilde{\Psi }$.
\begin{eqnarray}
\Psi &=&E-e_{5}cB,\quad \widetilde{\Psi }=E+e_{5}cB,\ E=(1/2)(\Psi +%
\widetilde{\Psi }),  \nonumber \\
e_{5}B &=&(-1/2c)(\Psi -\widetilde{\Psi }),\quad B=(1/2c)e_{5}(\Psi -%
\widetilde{\Psi }).  \label{cpe}
\end{eqnarray}
It is easy to find how the $F$ formulation and the $\Psi $ formulation are
connected $\Psi =(-1/c)(vF)$, $\widetilde{\Psi }=(-1/c)(\widetilde{F}%
v)=(1/c)(Fv),$ and conversely $F=(-1/c)(v\Psi )$, $\widetilde{F}=(-1/c)(%
\widetilde{\Psi }v).$ Then we can write ME (\ref{MEF}) in terms of $\Psi $
in a simple form
\begin{equation}
\partial (v\Psi )=-j/\varepsilon _{0},\quad \partial \cdot (v\Psi
)=-j/\varepsilon _{0},\ \partial \wedge (v\Psi )=0.  \label{mps}
\end{equation}
$\Psi $ is a mixed-grade multivector; the sum of an 1-vector and a 3-vector
(pseudovector). (\ref{mps}) can be written in $\left\{ e_{\mu }\right\} $
basis (with constant $e_{\mu }$) with the CBGQs as
\begin{equation}
((\Gamma ^{\alpha })_{\ \mu }^{\beta }\partial _{\alpha }\Psi ^{\mu
})e_{\beta }=-(j^{\beta }/\varepsilon _{0})e_{\beta },\quad (\Gamma ^{\alpha
})_{\ \mu }^{\beta }=\delta _{\quad \rho \nu }^{\alpha \beta }v^{\rho }g_{\
\mu }^{\nu }+e_{5}\varepsilon _{\quad \mu \nu }^{\alpha \beta }v^{\nu }.
\label{gps}
\end{equation}
where $\Psi ^{\alpha }e_{\alpha }=(E^{\alpha }-ce_{5}B^{\alpha })e_{\alpha }$%
. (\ref{gps}) contains both equations with $E$ and $B$ from (\ref{maeb}).
When $j^{\beta }=0$ the equation (\ref{gps}) becomes
\begin{equation}
((\Gamma ^{\alpha })_{\ \mu }^{\beta }\partial _{\alpha }\Psi ^{\mu
})e_{\beta }=0.  \label{gam1}
\end{equation}
We see from (\ref{gam1}) that ME for the free electromagnetic field become
Dirac-like relativistic wave equation for the free photon. But $\Psi $ is
real and there is no need for the probabilistic interpretation of $\Psi $!
This will be discussed elsewhere. $T(v)$ again can be written as a sum of
the $v-\parallel $ and the $v-\perp $ parts
\begin{equation}
T(v)=-(\varepsilon _{0}/4c)\left( \Psi \cdot \Psi +\widetilde{\Psi }\cdot
\widetilde{\Psi }\right) v+(\varepsilon _{0}/4c)\left( \Psi \cdot \Psi -%
\widetilde{\Psi }\cdot \widetilde{\Psi }\right) \cdot v.  \label{tps}
\end{equation}
The first term in (\ref{tps}) yields $U=-(\varepsilon _{0}/4)\left( \Psi
\cdot \Psi +\widetilde{\Psi }\cdot \widetilde{\Psi }\right) $ while the
second term gives $S=(\varepsilon _{0}/4)\left( \Psi \cdot \Psi -\widetilde{%
\Psi }\cdot \widetilde{\Psi }\right) \cdot v.$ The Lorentz force is given as
$K=(q/c^{2})\left[ u\cdot (v\cdot \Psi )+(u\cdot v)\Psi -v\wedge (u\cdot
\Psi )\right] .$

\textit{The complex} $\Psi =E-icB$ \textit{formulation}.- Sometimes it will
be useful to work with the complex $\Psi $; Clifford algebra is developed
over the field of the complex numbers. Then
\begin{equation}
\Psi =E-icB,\ \overline{\Psi }=E+icB,\quad E=(1/2)(\Psi +\overline{\Psi }),\
B=(i/2c)(\Psi -\overline{\Psi }),  \label{co}
\end{equation}
$\overline{\Psi }$ is the complex reverse of $\Psi $ and now it holds that $%
v\cdot \Psi =v\cdot \overline{\Psi }=0$. In contrast to the real $\Psi $ the
complex $\Psi $ is a homogeneous, grade-1, multivector. The $F$ and the
complex $\Psi $ formulations are connected as $\Psi =(1/c)F\cdot
v+(i/c)e_{5}(F\wedge v),$ and $F=(1/2c)(\Psi +\overline{\Psi })\wedge
v+(i/2c)\left( e_{5}(\Psi -\overline{\Psi })\right) \cdot v.$ Then the ME (%
\ref{MEF}) become
\begin{equation}
\partial \cdot (v\wedge \Psi )-ie_{5}\left[ \partial \wedge (v\wedge \Psi
)\right] =-j/\varepsilon _{0}.  \label{mecp}
\end{equation}
In $\left\{ e_{\mu }\right\} $ basis (with constant $e_{\mu }$) the ME (\ref
{mecp}) will have the same form as the equation (\ref{gps}) but in $(\Gamma
^{\alpha })_{\ \mu }^{\beta }$ $e_{5}$ is replaced by $i.$ Notice also that
the equations for the complex $\Psi $ are of the same form as the
corresponding equations written in the true tensor formulation with the
complex $\Psi $ in $\left[ 5\right] $ (or in the component form in the ''E''
coordinatization in $\left[ 7\right] $ and $\left[ 10\right] $). Again for $%
j^{\beta }=0$ we find that ME become Dirac-like relativistic wave equation
for the free photon. $T(v)$ can be expressed as a sum of the $v-\parallel $
part and the $v-\perp $ part as
\begin{equation}
T(v)=-(\varepsilon _{0}/2c)\left( \Psi \cdot \overline{\Psi }\right)
v-i(\varepsilon _{0}/2c)e_{5}\left[ \Psi \wedge \overline{\Psi }\wedge
v\right] .  \label{kot}
\end{equation}
Then $U=-(\varepsilon _{0}/2)\left( \Psi \cdot \overline{\Psi }\right) ,$
and $S=-i(\varepsilon _{0}/2)e_{5}\left[ \Psi \wedge \overline{\Psi }\wedge
v\right] .$

In summary, the approach to the relativistic electrodynamics presented in
this paper deals exclusively with geometric quantities, i.e., frame and
coordinatization independent quantities. All four formulations, with $F$, $E$
and $B$, real and complex $\Psi $ are equivalent and yield a complete and
consistent description of electromagnetic phenomena in terms of properly
defined quantities on the 4D spacetime. The formalism presented here will be
the basis for the relativistically correct (without reference frames)
formulation of quantum electrodynamics and, more generally, of the quantum
field theory. \bigskip

\noindent REFERENCES

\noindent $\left[ 1\right] $ D. Hestenes, \textit{Space-Time Algebra }%
(Gordon and Breach, New York, 1966); \textit{New Foundations for Classical
Mechanics }(Kluwer Academic Publishers, Dordrecht, 1999) 2nd. edn.

\noindent $\left[ 2\right] $ S. Gull, C. Doran, and A. Lasenby, in \textit{%
Clifford (Geometric) Algebras with Applications to Physics, Mathematics, and
Engineering,} W.E. Baylis, Ed. (Birkhauser, Boston, 1997), Chs. 6-8.

\noindent $\left[ 3\right] $ B. Jancewicz, \textit{Multivectors and Clifford
Algebra in Electrodynamics} (World Scientific, Singapore, 1989).

\noindent $\left[ 4\right] $ D. Hestenes and G. Sobczyk, \textit{Clifford
Algebra to Geometric Calculus }(Reidel, Dordrecht, 1984).

\noindent $\left[ 5\right] $ T. Ivezi\'{c}, Found. Phys. \textbf{8}, 1139
(2001).

\noindent $\left[ 6\right] $ T. Ivezi\'{c}, Found. Phys. Lett. \textbf{15},
27 (2002).

\noindent $\left[ 7\right] $ T. Ivezi\'{c}, Found. Phys. Lett. \textbf{12},
105 (1999); Found. Phys. Lett. \textbf{12}, 507 (1999).

\noindent $\left[ 8\right] $ A. Einstein, Ann. Physik. \textbf{17}, 891
(1905), tr. by W. Perrett and G.B.

Jeffery, in \textit{The principle of relativity} (Dover, New York).

\noindent $\left[ 9\right] $ H.N. N\'{u}\~{n}ez Y\'{e}pez, A.L. Salas Brito,
and C.A. Vargas, Revista Mexicana de F\'{i}sica \textbf{34}, 636 (1988).

\noindent $\left[ 10\right] $ S. Esposito, \textit{Found. Phys.} \textbf{28}%
, 231 (1998).

\end{document}